# Influence of surface chemistry on the electronic properties of graphene related carbon materials

Arunabhirm Chutia[a*], Fanica Cimpoesu[b], Hideyuki Tsuboi[c] and Akira Miyamoto[c,d]

[a]World Premier International (WPI) Research Center, Advanced Institute for Materials Research, Tohoku University, Sendai 980-8577, Japan

[b] Institute of Physical-Chemistry, Splaiul Independenței 202, Bucharest 060021, Romania

[c]Department of Applied Chemistry, Graduate School of Engineering, Tohoku University, sendai 980-8579, Japan

[d]New Industry Creation Hatchery Center, Tohoku University, Sendai 980-8579, Japan

*Corresponding authors:

Fax: +81-22-217-5954, E-mail address: arun@wpi-aimr.tohoku.ac.jp (Arunabhiram Chutia)

A theoretical study on the influence of organic functional groups on the electronic properties of graphene related carbon materials was carried out. Here we report, using density functional theory and tight-binding approach, that the best candidates for conducting supramolecular devices can be obtained by engineering the surface chemistry and stacking conformation of these materials.

## 1. Introduction

Graphene related carbon materials (GRCMs) have been extensively studied for last several decades due to their wide range of applications such as in battery electrodes, catalytic support, capacitors, gas storage and biomedical applications [i,ii]. The GRCMs can be broadly divided as pure and amorphous GRCMs. While, the chemical and physical properties of pure GRCMs such as graphite, fullerenes and carbon nanotubes have been thoroughly studied, the influence of physical and chemical features of amorphous GRCMs (a-GRCMs) such as carbon black, activated carbon, soot etc. on its electronic properties are not fully understood though, they have higher values of application [1,iii-ix].

Since early 1950s many workers carried out research to understand the relationship between the stability of a-GRCMs and their surface chemistry. For example, it was put forward that the stability of carbon black, an a-GRCM, are due to the presence of organic functional groups (OFGs) such as –OH, –CH, –COOH, –C=O etc. on the surface of its particles [x-xii]. FTIR spectral analysis by Juan et al showed that in the carbon black structures there are lots of aromatic C–C bonds and a large amount of nitrogen containing OFGs such as –NH, –CN, –CH, –OH and –COOH [xiii]. Pantea et al also reported the presence of carbon-sulfur groups [xiv,xv]. The XRD patterns of activated carbon, another a-GRCM, has revealed the presence of disordered graphite microcrystallites, with pores due to inter and intra crystallite units [ xvi,xvii]. In these materials the basic microcrystallite unit consists of 3-4 graphene layers and they have OFGs such as >C=O, –COOH, –C–N [xviii].

Another active area of research associated with a-GRCM is the study of the relationship between the surface chemistry of these materials and their physical properties such as electrical conductivity. In several papers it was reported that the electrical conductivity of a-GRCMs is dependent on the sample preparation procedures, on the surface area, structure etc [xix]. But in spite of numerous experimental studies on these materials only recently Pantea et al. employed modern spectroscopic techniques to investigate the relationship between electrical conductivity and surface chemistry of a-GRCMs i.e., carbon blacks. They reported different values of electrical conductivity for carbon black samples with similar surface chemistry. In another study Kumari et al showed that in certain amorphous carbon samples electrical conductivity varies with sulfur contents i.e. the electrical conductivity first increases with increasing sulfur content up to an optimum sulfur content then it drastically decreases [xx].

In addition to these experimental studies limited theoretical work on a-GRCMs has been reported e.g., the adsorption properties of water and phenols on soot and activated carbons respectively were reported by Efremenko et al and Hamad et al [vii,viii,xxi]. Kurita et al investigated the substitution of B, N and S in the graphene models to investigate the storage of Li atoms [xxii,xxiii]. Huang et al studied the structure and electronic properties of few layer graphenes by a first principle method [xxiv].

The above mentioned studies provided a wealth of information about a-GRCM but the effects of OFGs on electronic properties were not taken into account. Thus, in the present investigation we have taken into account of the influence of OFGs of a-GRCMs on their electronic and electrical properties. The electronic and electrical properties of functionalized graphenes have been investigated with reference to –OH and –SH groups.

## 2. Computational Details

Density functional theory (DFT) calculations were performed by DMol$^3$ program [xxv,xxvi]. All the calculations were performed at DNP level of basis set. Geometry optimization was performed using LDA/VWN exchange and correlation functionals.[xxvii] In order to improve the performance beyond LDA frame, the single point calculations were repeated with gradient corrected exchange-correlation functional, using the GGA/PBE ones [xxviii,xxix].

The electrical conductivity of the larger a-GRCM models were evaluated using a novel tight-binding quantum chemical molecular dynamics program, COLORS [xxx,xxxi]. In this algorithm the total energy for a given system is expressed by the following equation

$$E = \sum_{i=1}^{n} \frac{1}{2} m_i v_i^2 + \sum_{k=1}^{occ} \varepsilon_k + \sum_{i=1}^{n} \sum_{i=j+1}^{n} \frac{Z_i Z_j e^2}{R_{ij}} + \sum_{i=1}^{n} \sum_{j=i+1}^{n} E_{ij}^{rep}\left(R_{ij}\right) \qquad (1)$$

where the first, second, third and fourth terms represent the kinetic energy, the eigen values of all the occupied molecular orbital (orbital energy of valence electrons), the Coulombic interaction energy, and the exchange-repulsion interaction energy, respectively. Here, $m_i$ and $v_i$ represent the the mass and velocity of the atoms. The $Z_i$ and $Z_j$ stand for the atomic charges, while $e$ is the elementary electric charge and $R_{ij}$ is the internuclear distance. The exchange-repulsion term $E_{rep}\,(R_{ij})$ is given by

$$E_{rep}(R_{ij}) = b_{ij} \exp\left[\frac{(a_{ij} - R_{ij})}{b_{ij}}\right] \qquad (2)$$

where, the parameters $a$ and $b$ represent the sum of the size and stiffness of atoms, $i$ and $j$. The cutoff value in the exchange repulsion force is determined as the distance of the half of the shortest cell size length.

A novel conductivity simulator based on Monte Carlo method was developed to estimate the electrical conductivity of the systems. First a single point tight-binding calculation was performed to obtain the energies and eigenvectors of the molecular orbital model. Then, the carrier mobility was estimated by means of Monte Carlo simulation. In this procedure the number of carriers, $n$, is assumed to follow the Fermi distribution. The conductivity σ is expressed by the equation

$$\sigma = ne\mu \quad (3)$$

where, $n$ is number of carriers and $\mu$ corresponds to the mobility. A detailed description of the methodology can be found elsewhere [xxxii].

The classical molecular dynamics simulations were carried using Ryudo program [xxxiii]. The MD simulations were done using NPT ensemble, with pressure (P) of 1 atm, temperature (T) of 298 K, with 0.1 fs time step for 450000 and 550000 time steps.

## 3. Results and Discussion

### *3.1 Models*

In order to investigate the influence of OFGs on the electronic properties of nano-graphene clusters functionalized with –OH and –SH groups, we considered here different substitution patterns, as compared to the unsubstituted graphene sheets saturated with H. Figure 1 (a) shows the hydrocarbon skeleton with ideal $D_{6h}$ symmetry. Figures 1 (b) and (c) are showing the models substituted with six –OH groups and six –SH groups in the periphery. These systems are referred as G(OH) and G(SH) where G stands for pristine graphene. Figure 1 (d) and (e) show the models with mixed substitution, with both –OH and –SH groups. The (1d) case shows a 1,3,5-like distribution of chemically equivalent groups, having a $C_{3h}$ molecular symmetry (referred as G(OS-$C_3$h). The (1f) has a 1,2,3, like substitution, having the lower $C_s$ symmetry. (labeled and G(OS-$C_s$).

Experimental studies revealed that graphene sheets in a-GRCMs such as carbon blacks and activated carbons are arranged in stacked 4-5 layers and the number of stacked layers can decrease with increasing amorphicity of a-GRCMs samples [ix,xv]. Thus, in the present investigation single, double, triple and quadruple stacked nano-graphene models were considered (e.g., see Figure 2 for G(OH) models). To mimic the experimentally observed interlayer distances the basic structural units (BSUs), model geometries with stacking restricted to the inter-plane separations of 3.35 Å were considered. It is well known that the BSUs in a-GRCMs e.g., carbon blacks are turbostratic i.e., stacking of graphitic monolayer have no periodicity along *c* axis. However, for the sake of simplicity we started with systems with ideal initial order letting after that molecular dynamics to induce a thermal disordering. As shown in Figure 3 we considered two patterns of stacking: ABA (C atoms in alternative layers directly on top of each other) and AAA (C atoms in consecutive layers directly on top of each other) [xxxiv]. The ABA type is known to be present in natural graphite, these systems being also submitted to the full DFT optimization. The AAA pattern is a hypothetical one, considered here for comparison.

Large models with G(OH) basic structural units were submitted to molecular dynamic simulations. The initial geometry was randomly generated as shown Figure 4 (a) with periodic boundary condition. The final structures, after the completion of MD simulation had densities of 1.11 g/cm$^3$ and 1.63 g/cm$^3$ respectively [Figure 4 (b), (c)] with 468 atoms in both the models. These models were further employed to investigate the influence on electrical properties due to morphological changes caused by compression pressure as reported in literature [xxxv].

### *3.2. Electronic properties of constrained functionalized models*

A general characterization of the electron structure was done analyzing the regularities recorded in the highest occupied molecular orbital (HOMO), lowest unoccupied molecular orbital (LUMO) energies and HOMO-LUMO gap ($E_g$), along the series of the AAA and ABA stacked G(OH) and G(SH) molecules.

As shown in Figure 5 the $E_g$ of all the systems generally decreased with the number of layers. The plateau tendency appearing from *n*=3 to the 4 series suggest that the Eg values for larger systems will remain approximately close to those reached in the *n*=4 case. As shown in Table 1 and Figure 6 the energy gaps in AAA stacking case are lower than those of the ABA situation. This suggests that the ABA systems are more stable, in line with their occurrence in experimental evidences or ab initio optimized geometries. The $E_g$ as function of stacking pattern is mainly determined by the nature of interlayer $\psi_{2p}^{(z)}(r)$ - $\psi_{2p}^{(z)}(r)$ interaction. The $\psi_{2p}^{(z)}(r)$ components are generating the $\pi$ orbitals in the plane of the graphene sheet [xxxiv]. In the AAA case (see Figure 7) there is a maximized overlap between aligned $\psi_{2p}^{(z)}(r)$type orbitals on closer atoms from adjacent sheets. By contrary, in the ABA stacking patterns a weaker side-on overlap between the $\psi_{2p}^{(z)}(r)$ orbitals occurs. Apparently the AAA orbital factors are favoring the inter-planar interactions. However this enters in competition with the intra-planar bonding since it works in the sense of decoupling the established π bonds in the aromatic systems. At the same time the AAA pattern is destabilized by the repulsion (electrostatic and more complex effects that can be called Pauli repulsion) occurring between π clouds of the neighbor aromatic molecular planes. This repulsion effect determines the fact that the AAA is not the preferred stacking. The above picture can also be understood in the terms of the maximum hardness principle that states that the most stable systems are correlated with higher chemical hardness and therefore with larger HOMO-LUMO gaps.

Further insight in the electron structure is done analyzing the partial density of states (PDOS) components of density of states (DOS) of these systems. For this purpose the G(SH) system was considered, introducing supplementary computational experiments regarding the dependence of PDOS diagrams on the interplanar distance. Thus the system G(SH) with the regular spacing 3.35 Å and AAA stacking was taken as shown in Figure 6 (a) in comparison to the case with the slightly enhanced departure, 3.50 Å, labeled as G(SH)' (Figure 8 (b) ). At larger inter-planlar separation the $E_g$ gap was lowered, from 0.544 eV to 0.125 eV proving that this quantity is modulated by the interlayer overlap effects. The dominant contribution in PDOS curves comes from the valence shell of the carbon main components. The valcence shell of sulphur atoms (S3p) is also mixed up in many orbitals including the frontier ones, with relatively small contribution, corresponding to their lower chemical ratio in the structure.

The shrinking of the gap occured mainly by the lower shift of the LUMO energy from -2.672 eV in G(SH) to -3.169 eV in the G(SH)′, while HOMO approximately kept its peak. The case of OH substituted graphenes was followed drawing the dependence of HOMO and LUMO levels along the series depending on the number of layers n, and stacking manner (Figure 6). At the trivial situation $n$=1 the AAA and ABA obviously coincided since there was only one molecular component and no stacking involved. The AAA series showed a strong dependence with the number of $n$ elements, showing a quick tendency to smaller gaps by the opposite variation HOMO that rises its level and LUMO that descends in energy scale. In the ABA case the gap variation is less pronounced because of almost parallel evolution of HOMO and LUMO that showed a slight increase in energy as function of the *n* number. This showed that in the AAA case there was a rather drastical dependence on the interlayer effect due to the direct overlap effects in the distant C...C interaction.

The HOMO–LUMO energy gaps can be correlated with electrical conductivity. This is due to the fact that the frontier orbital energy gap can be roughly assigned to the so called chemical hardness, $\eta \sim -\frac{\varepsilon_{HOMO} - \varepsilon_{LUMO}}{2}$. The chemical hardness correlates directly to the ‘‘rigidity’’ or resistance of electron clouds on molecules to deformation against given perturbations. The low hardness, which correlates to small HOMO–LUMO energy gaps, corresponds to easily deformable electron cloud. The inverse of hardness, $S = \frac{1}{\eta}$, known as chemical softness, is proportional to the magnitude of electrical conductivity. A plot of chemical softness versus numbers of layers for AAA stacked models of G(OH), G(SH), G(OS-$C_{3h}$) and G(OS-$C_s$) is shown in Figure 9. It was seen that for the low dimensional systems, $n$=1,2 and the softer systems are the G(SH) ones, in line with the well known scales of chemical hardness that rate sulphur as soft and oxygen as hard components. The fact that GS(OH) n=3,4 systems progressively reveal as systems with highest global softness, was an interesting finding of our numerical experiments on the AAA stacked systems. This could be interpreted by the fact that electron density injected from oxygen into the graphene planes disfavoring (by increased repulsion) the interlayer bonding, enhancing the mobility of the electronic cloud. The sulphur G(SH) show lower effects of this type because of lower electronegativity of the sulfur and because of the fact that S atoms are able to exert opposite effects like the π-witdrawing by the action of its virtual *d*-type AOs. Finaly, the mixed subsistuent systems showed the lowest softness (not an intermediate one as possibly expected) because of the fact that the electron localization effects, driven by the electronegativity difference of OH and SH groups traps a part of the electron density at the periphery of the stacked graphenes. This lowered the global softness and potential conductivity. The computational experiments illustrated with changing the chemical constitution and the stacking conformation encouraged the theoretical search for best chemical candidates, in aiming for conducting supramolecular devices. It is suggested that these electric and electronic properties could be engineered by tuning the OFGs of the models.

## 4. The Optimized ABA functionalized graphene models

In the above sections investigation of electronic properties of the systems were carried out on constrained graphene models functionalized with –OH and –SH OFGs. These models provided information about the fundamental parameters in molecular and supramolecular constitution of functionalized GRCMs i.e., and the correlation with electronic and electrical properties. Now we bring the computational experiment at a new level, considering the full quantum chemical optimization of the molecular geometries.

The ABA stacked models were energetically more stable as compared to the AAA patterns thus, we performed optimization on the ABA systems from the smaller two-layered model with 156 atoms up to the four layered case with 312 atoms.

### *4.1. Geometry*

In the initial structures of all the G(OH) and G(SH) models, like bulk graphite, interlayer distance was set to 3.35 Å. As shown in Table 2 after optimization of the structures for bi, tri and tetra layers of G(OH) and G(SH) models the interlayer distance decreased. The decrease in the interlayer distance for the G(OH) models is larger than that of the G(SH) models. This is attributed to the hydrogen bonding between the BSUs in G(OH). The G(SH) models also exhibited similar long range effects, but the hydrogen bonding due to the –SH groups is weaker than those of the –OH ones [xxxvi]. The layered components of the stacked systems showed slight deviation from their planarity. But the deviation from planarity was a bit more pronounced for the G(OH) models than for the G(SH) ones, proving that the strength of the hydrogen bond is a factor that may cause, as side effect the recorded plane puckering. The strong hydrogen bonding between the atoms from different planes in G(OH) model lead to the deviation from planarity. Indeed, it was seen that (see Figure 9) the regions where the distance between the –O---HO– of two different BSUs are ~1.9 Å the geometries of graphene monolayers had significantly deviated from their planarity in the peripheral regions.

In the case of G(SH) models the shortest distance between the interacting atoms in different planes was ~2.3 Å corresponding to a long hydrogen bond. A particularity of the –S---HS– bonds is the fact that some of the CSH angles were more perpendicular (See Figure 10) than the COH angles due to –O---HO– bonds. This was related with the different hybridization abilities of oxygen and sulfur atoms. Earlier works on multipolar electrostatic treatment of H bonds revealed that the O---H interaction is dominated by charge-charge attraction, while S---H interaction is due to the charge (H) – quadrupole (S) interactions [xxxvi]. For the formation of H bonds to oxygen the monopole-monopole interaction prefers a linear orientation, while the monopole-dipole and monopole-quadrupole interaction prefers a perpendicular orientation [xxxvi,xxxvii]. The establishing of the optimal geometry by hydrogen bonds is in competition with the optimal placement for pure π-π stacking due to the carbon skeleton. Since the S---H interactions are less stronger than the O...H ones these could enforce only lesser deviation from the planarity than in the case of G(OH) models. In addition to it due to the higher perpendicularity of the –CSH angle the geometrical deviation to release steric strain caused by the proximity of –SH groups to H atoms was also reduced [xxxviii].

### *4.2. Electronic Properties*

In the following we report the calculated frontier orbital parameters, HOMO, LUMO energies, and their $E_g$, considered at the optimized geometries of the ABA stacking pattern. From Table 3 it was seen that these systems followed similar decreasing trend with the increase in the number of stacked layers, similar to our discussion on the constrained geometry molecules. The $E_g$ for all the ABA models of G(OH) and G(SH) were higher than their ABA unoptimized counterparts. This is a direct expression of the orbital stability gained after the geometry optimization.

Figure 11 shows the changes in the energies of the molecular orbitals from (HOMO-5) to (LUMO+5) of G(OH) ABA models constituting of two layers with and without optimization. For further clarity all the energies of [(HOMO-5) to (LUMO+5)] have been summarized in Table 4. With the optimization, due to the reduced distances between the two π-electron rich beds, the electrostatic repulsion caused the shifting of the molecular orbital to higher energies. As shown in Figure 12, on comparing the PDOS of the optimized and unoptimized G(OH) models with two ABA stacked functionalized graphene layers it was seen that for the unoptimized structures the signatures due to C *2p* orbitals are comparatively broader near the highest occupied molecular orbital reflecting stronger $\pi - \pi$ coupling.

Table 3 also showed the chemical hardness and chemical softness of all these systems having similar trends as those of the constrained geometries for ABA G(OH) models.

**5. Electrical Properties of Functionalized Graphenes at Different Densities**

In the above sections, the influence of OFGs such as –OH and –SH on the electronic and electrical properties of GRCMs was investigated. It was seen that not only the OFGs but the interaction between the π-electron rich beds also influence the electronic properties. It was shown that by engineering the interlayer spacing between the BSUs in GRCMs the electronic and electrical properties could be also altered. As an extension to this work investigation on models with different densities generated by Molecular dynamics (MD) simulations was carried out (See Figure 4). This investigation was carried out keeping in mind that different forms of GRCMs have different densities and it was reported that carbon black GRCMs with different densities have different conductivities [xxxv].

On the MD generated models the electrical conductivity was estimated using a novel tight-binding program "Colors". The interlayer spacing between the BSUs of both structures obtained after MD simulation and calculated electrical conductivity, are summarized in Table 5. It was clearly seen from Table 5 that the electrical conductivity of the functionalized models were enhanced by the increase in density of the models. This increase in electrical conductivity was attributed to the increase in the electrical contacts between the BSUs. Thus the electronic and electrical properties can be engineered by changing the densities of the carbon materials at the molecular level.

## 6. Conclusions

A theoretical investigation on the influence of OFGs on the electrical properties of GRCMs was carried out. It was illustrated that by changing the chemical constitution and the stacking conformation the best chemical candidates for conducting supramolecular devices could be obtained.

It was seen that in the layered graphene models though the electronic properties such as energy gap showed a decreasing trend with increasing numbers of stacked layers it was dependent on the stacking patterns of the graphenes and on the interactions between them. For the oxygen modified graphenes the electronic properties were seen to be independent of the stacking patterns of the graphenes due to the electron withdrawing nature of oxygen atoms. However, in the sulfur modified graphenes the electronic and electrical properties are dependent on the stacking patterns. The reason for the anomalous behavior of sulfur-modified graphenes is understood in terms of the interaction between the π-electron rich beds. It was seen that by engineering the interlayer distances between the graphene layers the electronic and henceforth the electrical properties of GRCMs could be altered. Estimation of electrical conductivity, using a novel tight-binding code – COLORS, for larger models at different densities also revealed that electrical conductivity of the functionalized graphenes is dependent on the interlayer distances.

**Acknowledgements**

AC takes the pleasure to thank Prof. Michio Tokuyama, Professor of Condensed Matter Physics, WPI-AIMR, Tohoku University, for his encouragements during the preparation of this manuscript.

Table 1. HOMO, LUMO energies and energy gaps in eV of AAA and ABA stacked carbon black models.

| No. | CB(OH) | | | | | | CB(SH) | | | | | |
|---|---|---|---|---|---|---|---|---|---|---|---|---|
| | AAA | | | ABA | | | AAA | | | ABA | | |
| Layers | HOMO | LUMO | Eg | HOMO | LUMO | Eg | HOMO | LUMO | Eg | HOMO | LUMO | Eg |
| 1 | -4.295 | -2.397 | 1.898 | -4.295 | -2.397 | 1.898 | -4.714 | -2.906 | 1.808 | -4.714 | -2.906 | 1.808 |
| 2 | -3.663 | -3.022 | 0.641 | -4.168 | -2.976 | 1.192 | -3.648 | -3.27 | 0.378 | -3.963 | -2.967 | 0.996 |
| 3 | -3.326 | -3.004 | 0.322 | -3.940 | -2.838 | 1.102 | -3.252 | -2.708 | 0.544 | -3.734 | -2.833 | 0.901 |
| 4 | -3.303 | -3.143 | 0.160 | -3.764 | -2.748 | 1.016 | -3.219 | -3.05 | 0.169 | -3.602 | -2.728 | 0.874 |

Table 2. Interlayer distances of the G(OH) and G(SH) models

| No. of layers | Interlayer distance (Å) | |
|---|---|---|
| | G(OH) | G(SH) |
| 1 | 0 | 0 |
| 2 | 3.266 | 3.274 |
| 3 | 3.213 | 3.274 |
| 4 | 3.193 | 3.252 |

Table 3. HOMO, LUMO energies (eV), HOMO-LUMO energy gap (eV), Chemical Hardness (η) and Chemical Softness (S) of G(OH) and G(SH) ABA optimized models

| | Optimized G(OH) | | | | | Optimized G(SH) | | | | |
|---|---|---|---|---|---|---|---|---|---|---|
| Layers | HOMO | LUMO | Eg | η | S | HOMO | LUMO | Eg | η | S |
| 1 | -4.295 | -2.397 | 1.898 | 0.949 | 1.054 | -4.714 | -2.906 | 1.808 | 0.904 | 1.106 |
| 2 | -4.043 | -2.470 | 1.573 | 0.787 | 1.271 | -4.477 | -2.951 | 1.526 | 0.763 | 1.311 |
| 3 | -3.764 | -2.334 | 1.430 | 0.715 | 1.399 | -4.357 | -2.952 | 1.405 | 0.703 | 1.423 |
| 4 | -3.683 | -2.313 | 1.370 | 0.685 | 1.460 | -4.368 | -3.048 | 1.320 | 0.660 | 1.515 |

Table 4. Energy of (HOMO-5) – (LUMO+5) orbitals of optimized and unoptimized ABA models of G(OH) [number of layers = 2]

| Orbitals | G(OH) | |
|---|---|---|
| | Optimized | Unoptimized |
| HOMO-5 | -4.712 | -4.717 |
| HOMO-4 | -4.654 | -4.67 |
| HOMO-3 | -4.358 | -4.407 |
| HOMO-2 | -4.318 | -4.337 |
| HOMO-1 | -4.046 | -4.198 |
| **HOMO** | **-4.043** | **-4.168** |
| **LUMO** | **-2.470** | **-2.976** |
| LUMO+1 | -2.445 | -2.967 |
| LUMO+2 | -2.294 | -2.824 |
| LUMO+3 | -2.193 | -2.78 |
| LUMO+4 | -1.613 | -2.332 |
| LUMO+5 | -1.609 | -2.296 |

Table 5. Density, interlayer spacing of the final structures obtained after MD simulation and estimated electrical conductivity using tight-binding approach.

| System | Density ($g/cm^3$) | Interlayer spacing (Å) [Average] | Electrical conductivity (S/cm) |
|---|---|---|---|
| G(OH) | 1.11 | 3.97 | 3.54 |
| | 1.64 | 3.73 | 5.53 |

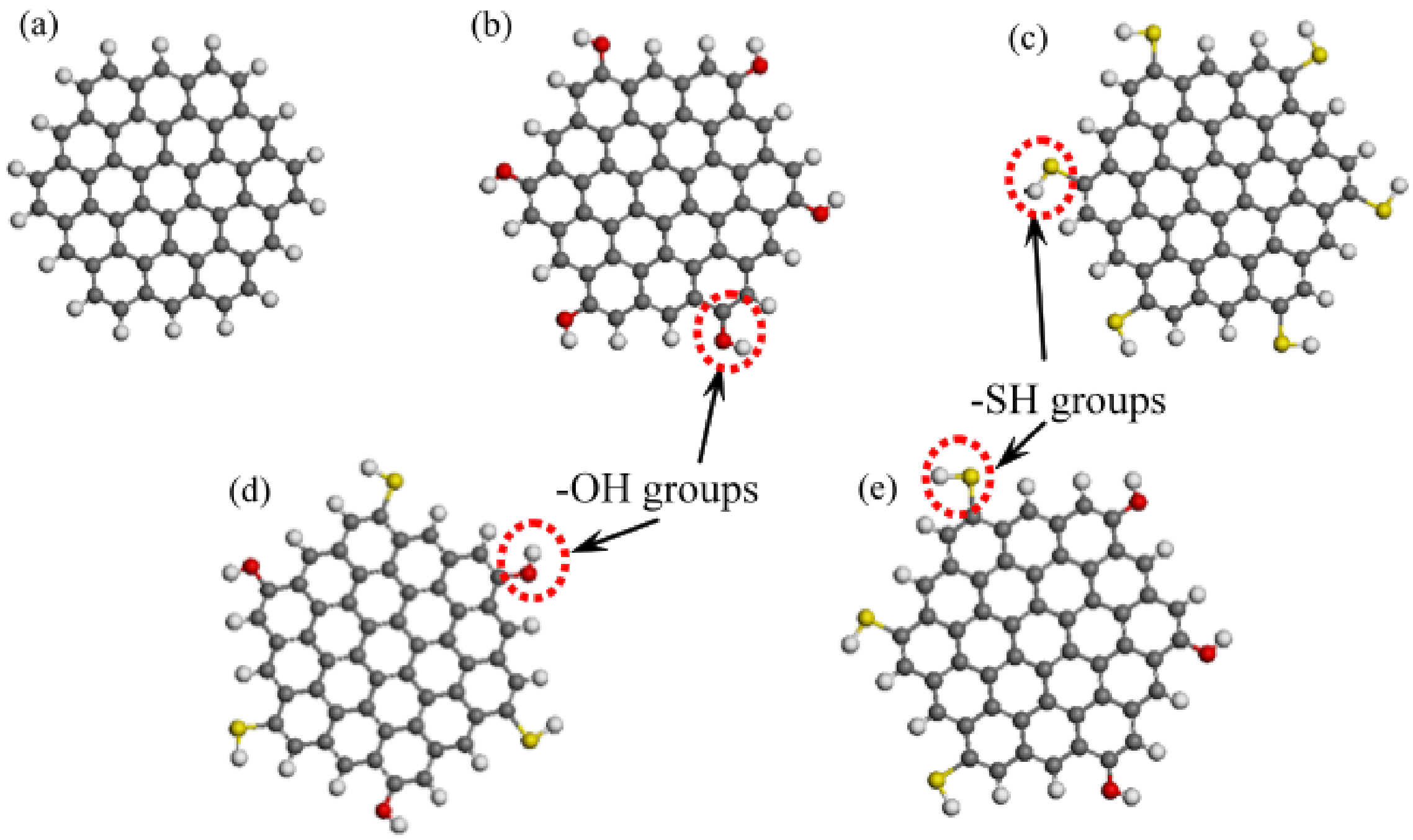

**Figure 1** (a) Pristine graphite model saturated with hydrogen atoms, (b) G(OH), (c) G(SH), (d) G(OS-$C_{3h}$) and (e) G(OS-$C_s$).

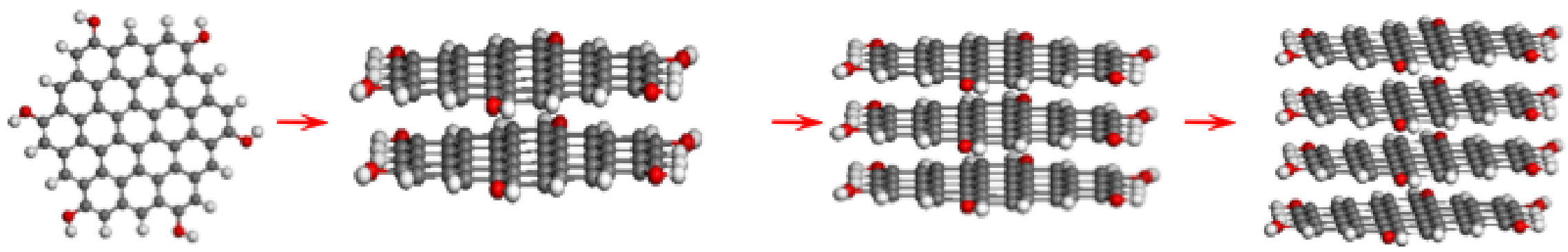

**Figure 2** GO(OH) models with (a) one, (b) two, (c) three and (d) four layers in AAA attacking patterns with an interlayer distance of 3.35 Å.

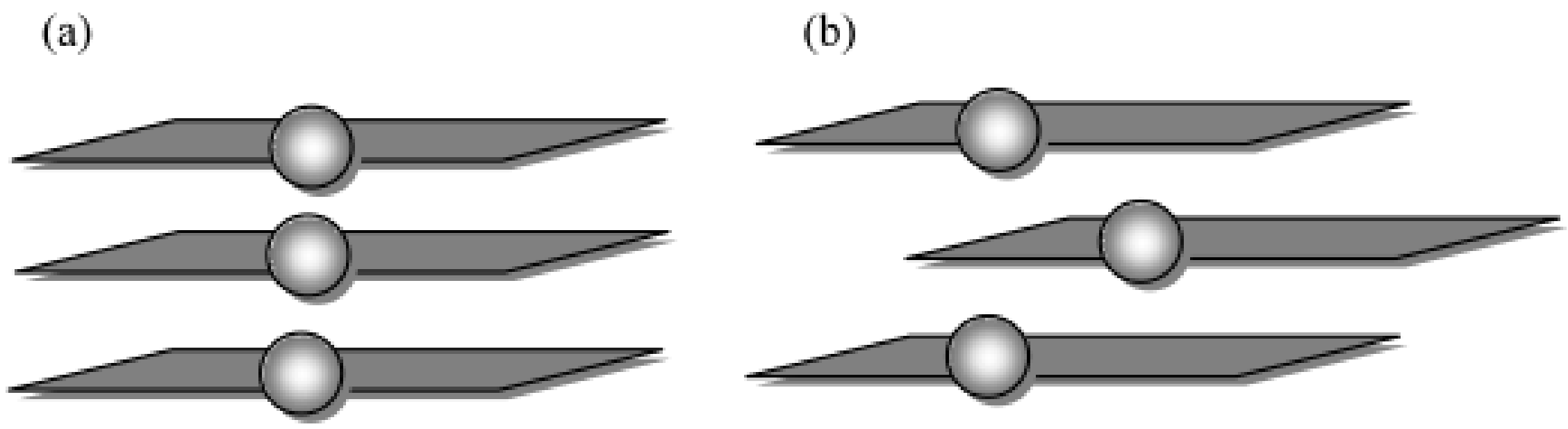

**Figure 3** Schematic representations of (a) AAA and (b) ABA stacking patterns

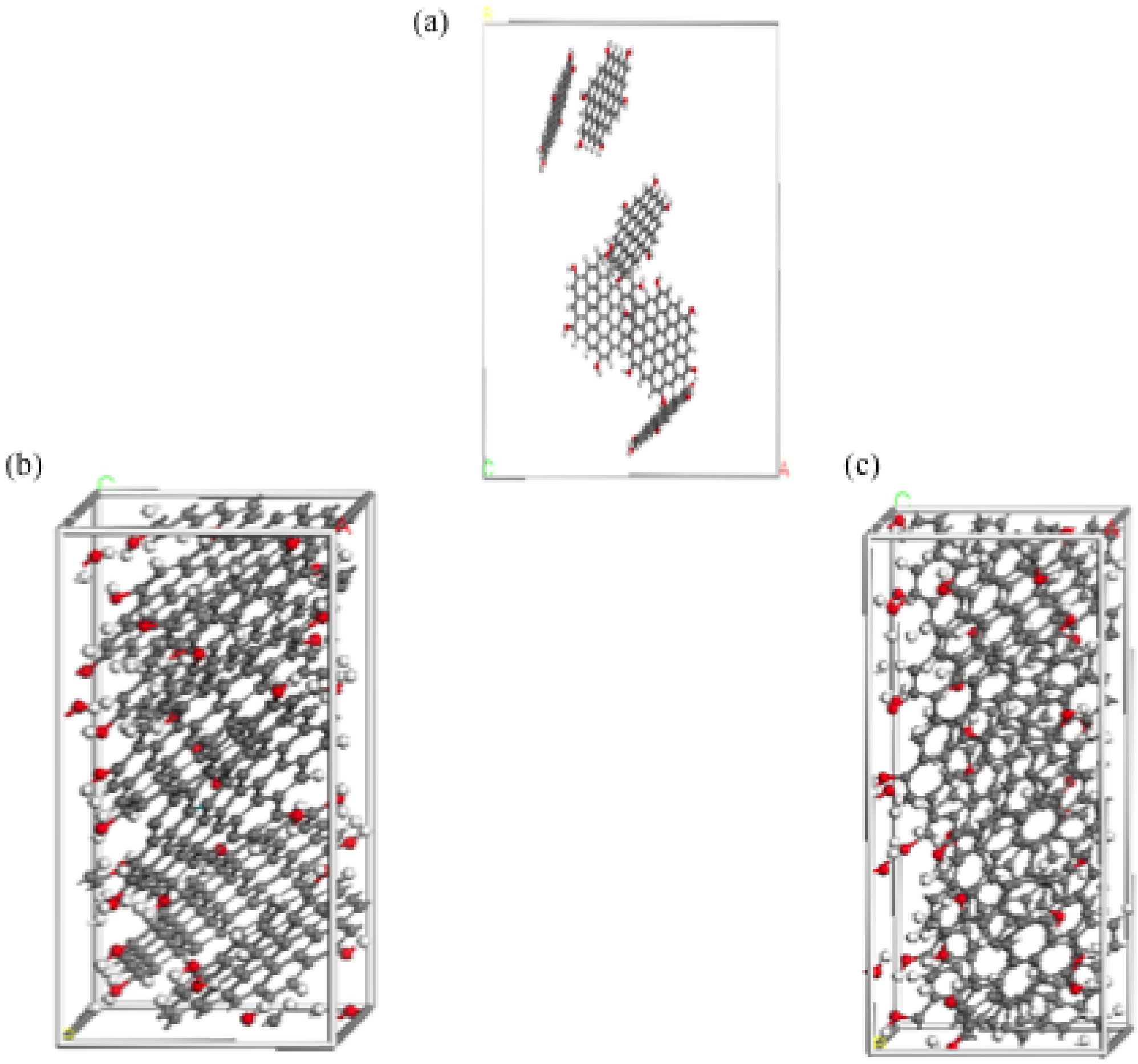

Figure 4 (a) Initial model of graphene related carbon materials to generate models with densities of (b) 1.11 g/cm$^3$ and (b) 1.64 g/cm$^3$.

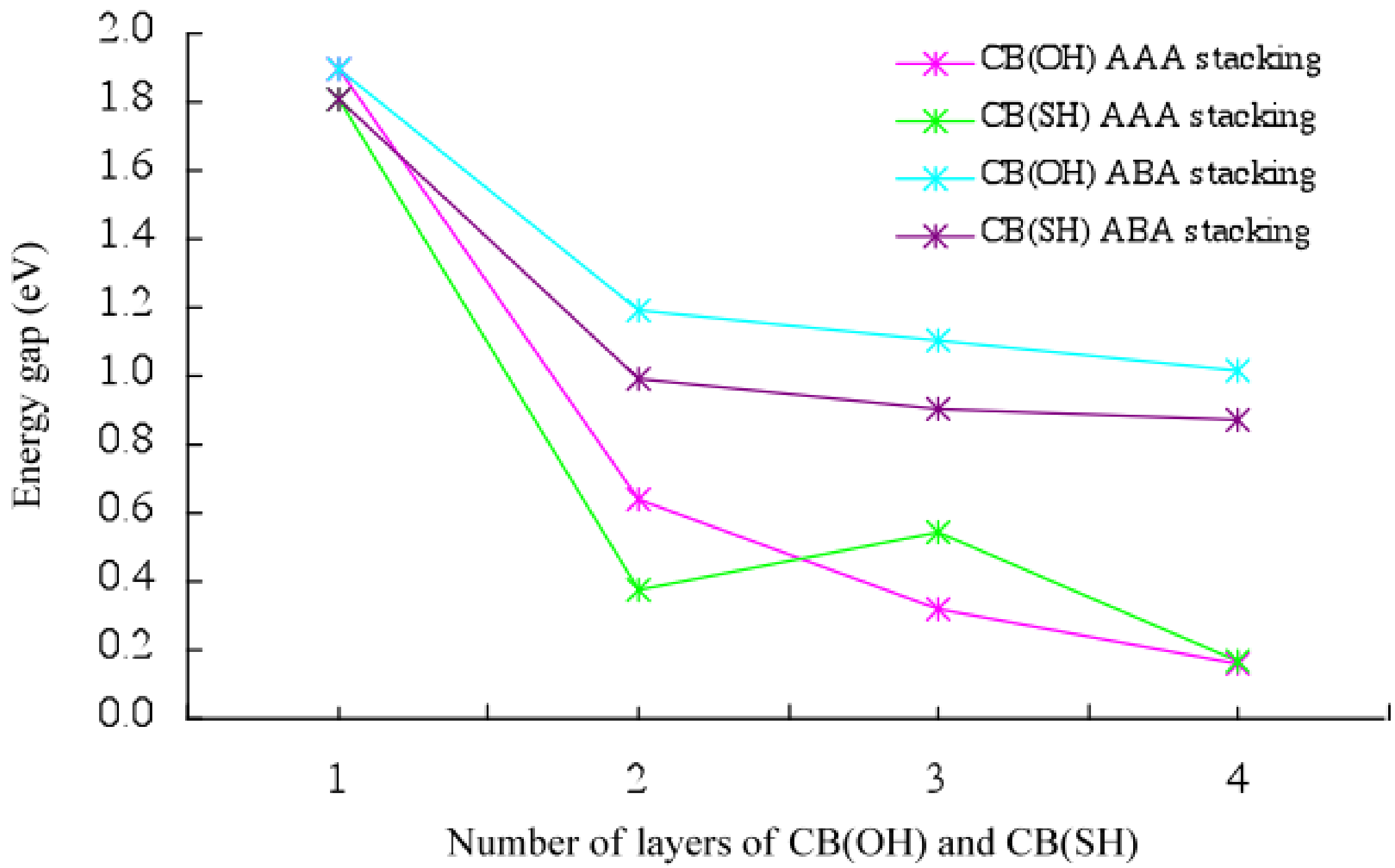

**Figure 5** Energy gap (eV) versus number of layers of CB(OH) and CB(SH) in AAA and ABA stacking patterns.

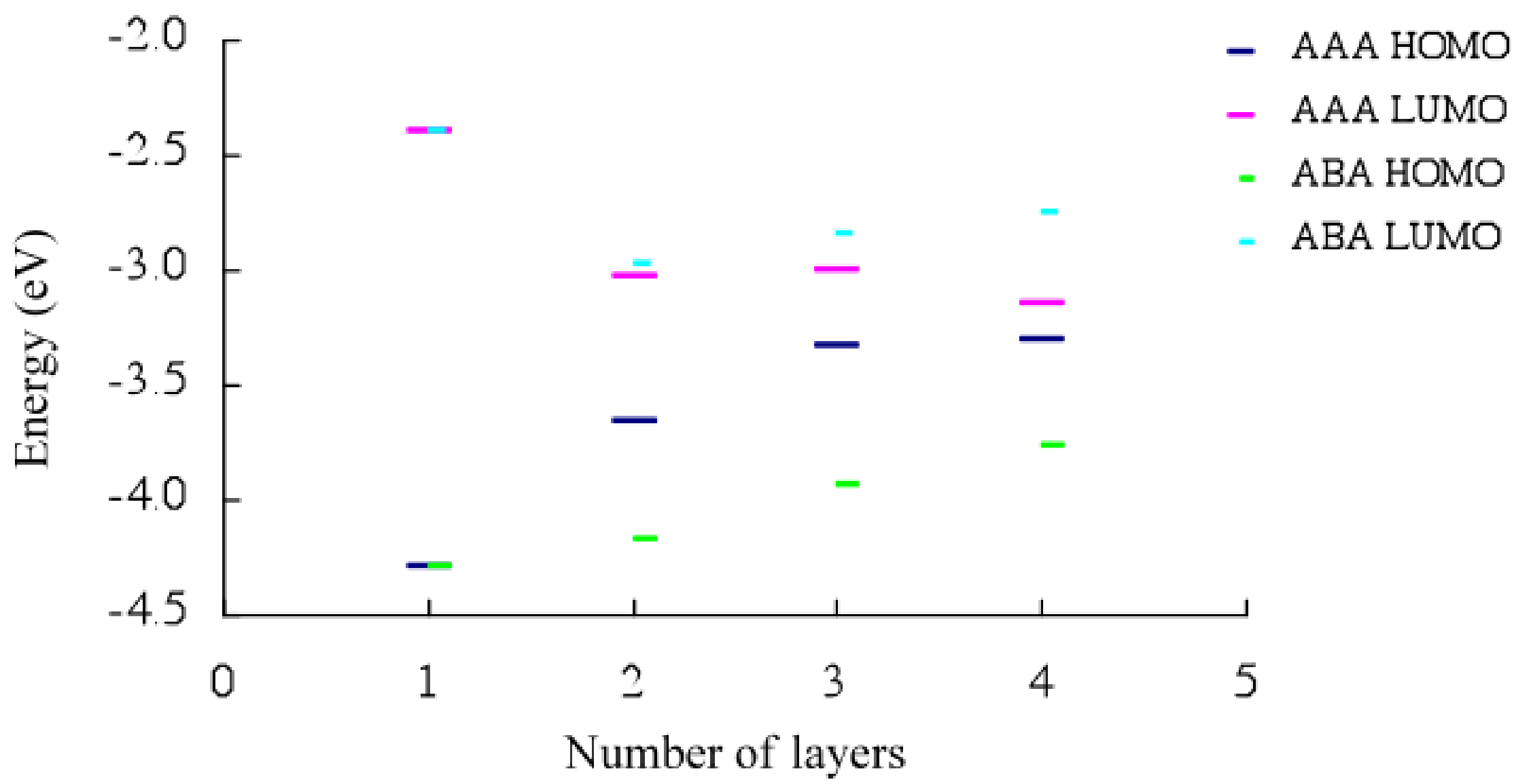

**Figure 6** HOMO and LUMO of AAA and ABA stacked G(OH) models

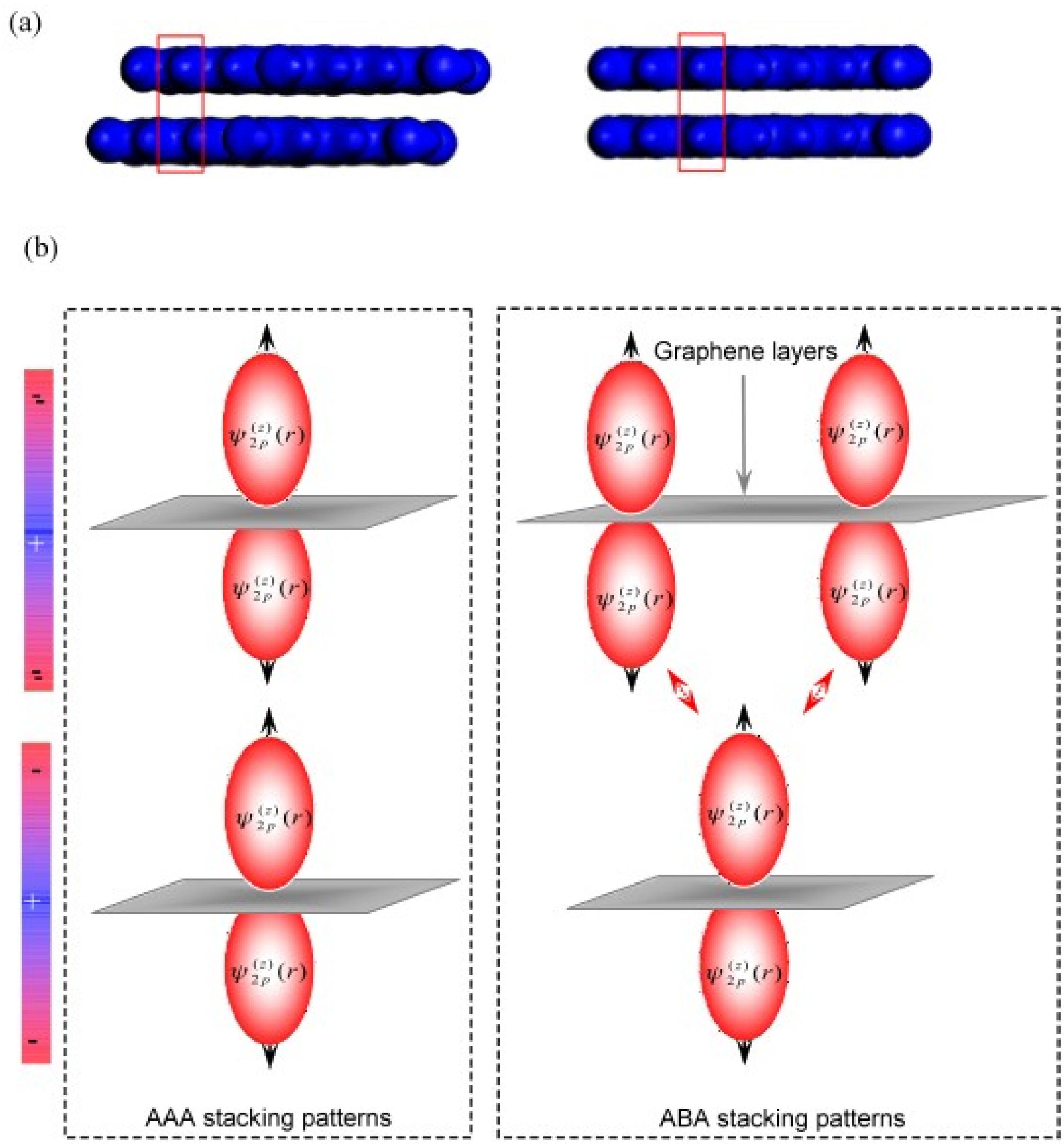

**Figure 7** A schematic representation of orbital interaction in (a) AAA and (b) ABA stacking patterns

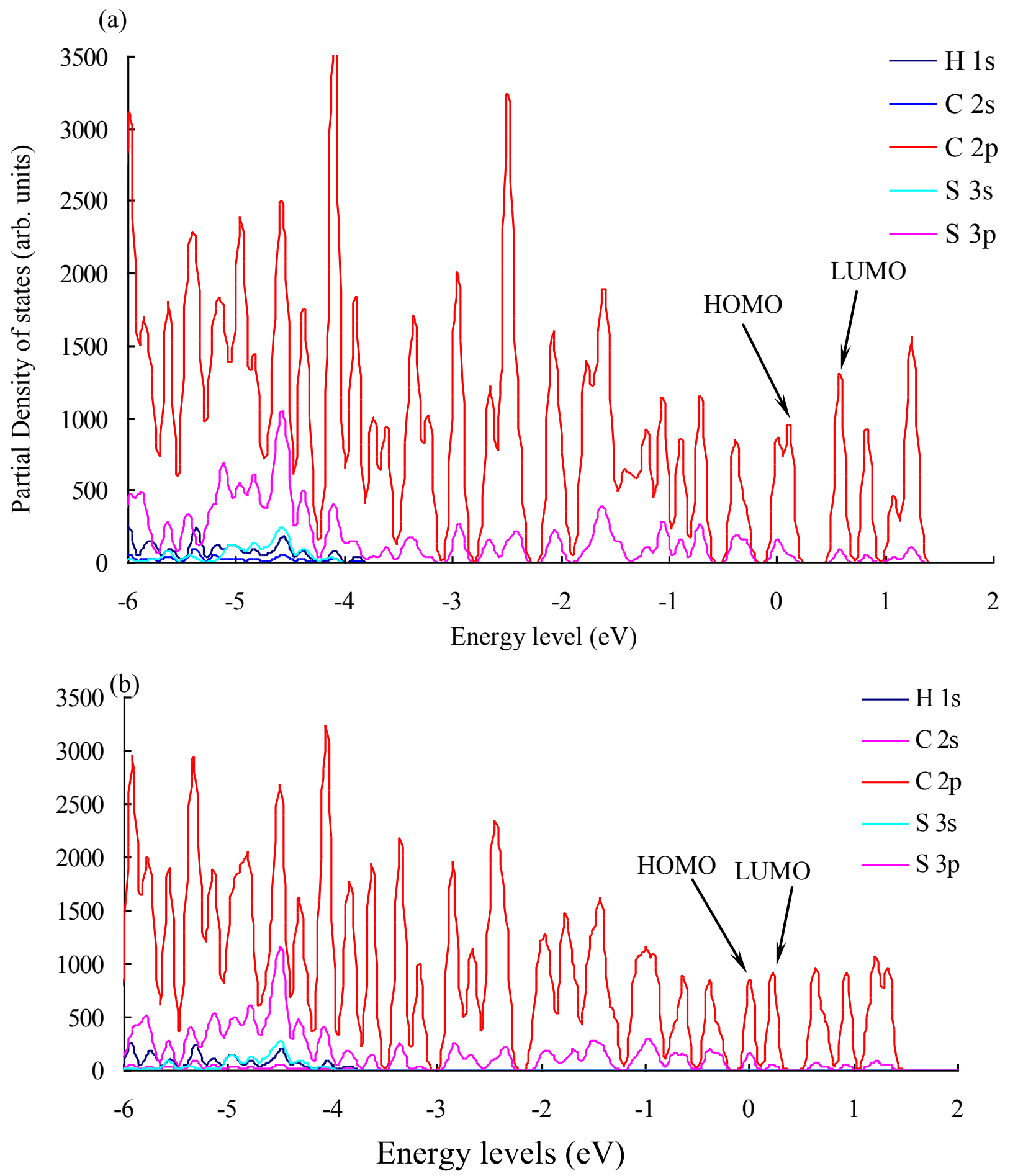

**Figure 8** Partial density of states for three layered AAA staked (a) CB(SH) and (b) CB(SH) ´

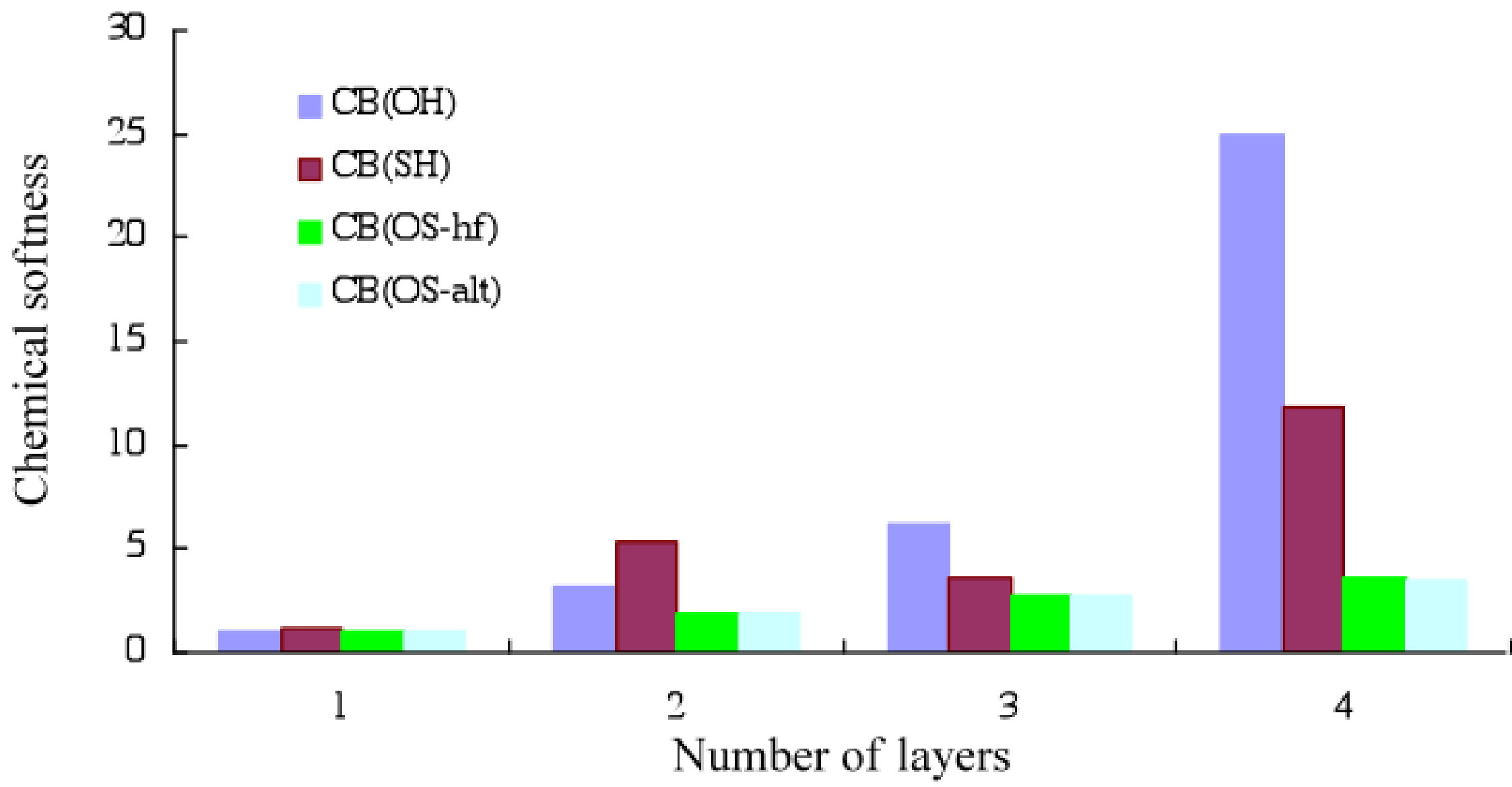

**Figure 9** Chemical softness versus numbers of layers for AAA stacked models of CB(OH), CB(SH), CB(OS-hf) and CB(OS-alt)

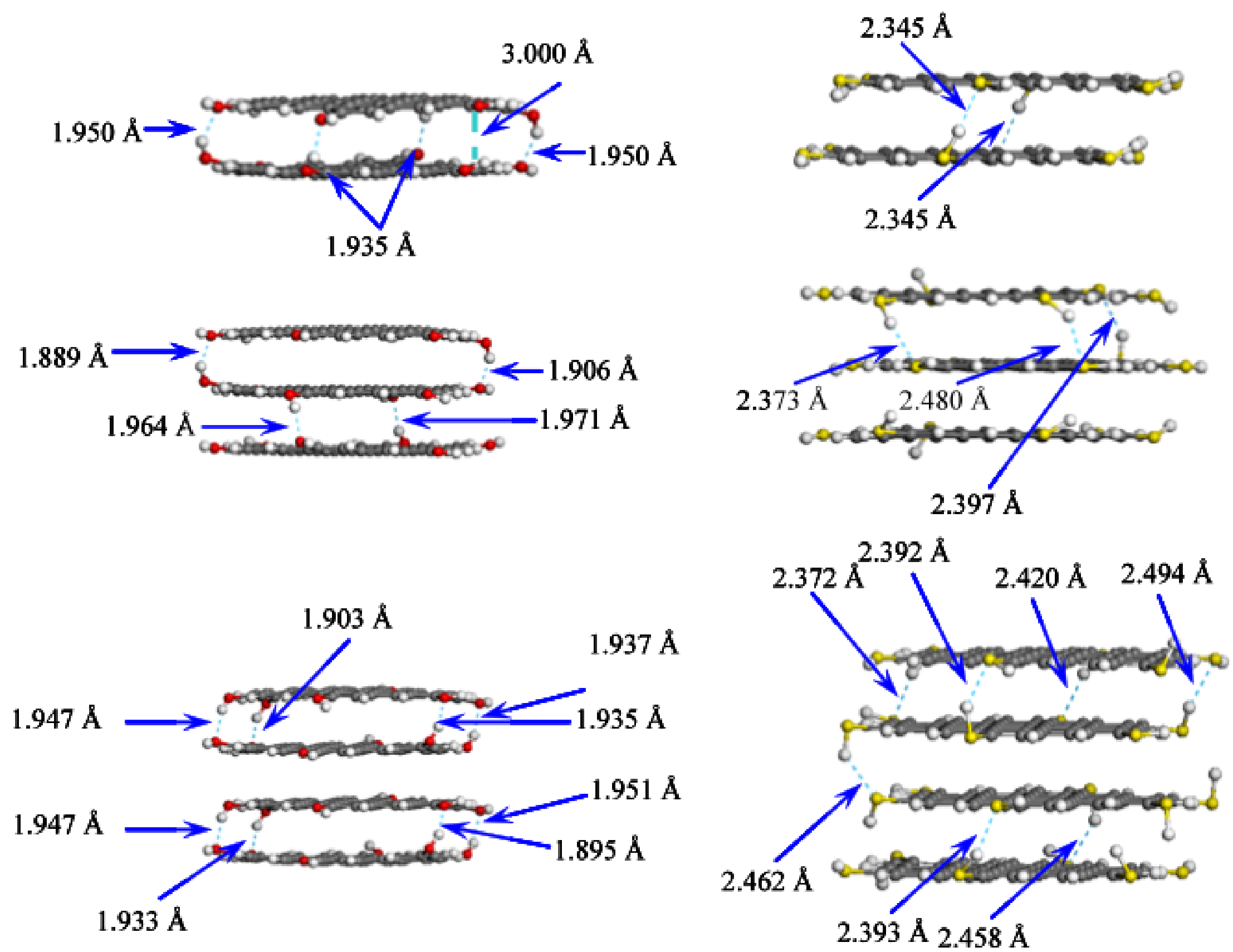

**Figure 10** Optimized geometries of G(OH) and G(SH) models

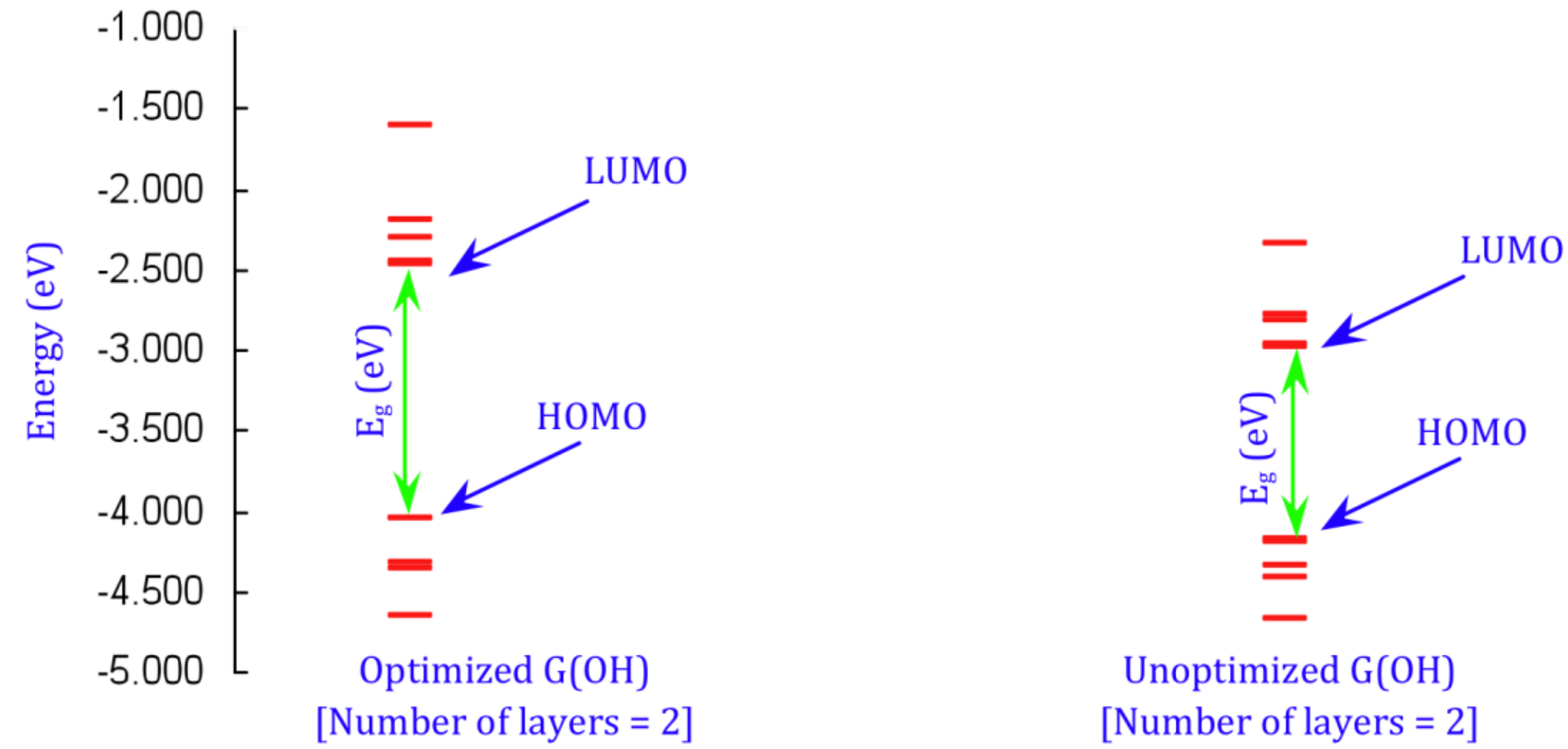

Figure 11 Energy level diagrams for (HOMO-5) – (LUMO+5) of (a) optimized (b) unoptimized ABA models of G(OH) [number of layers = 2]

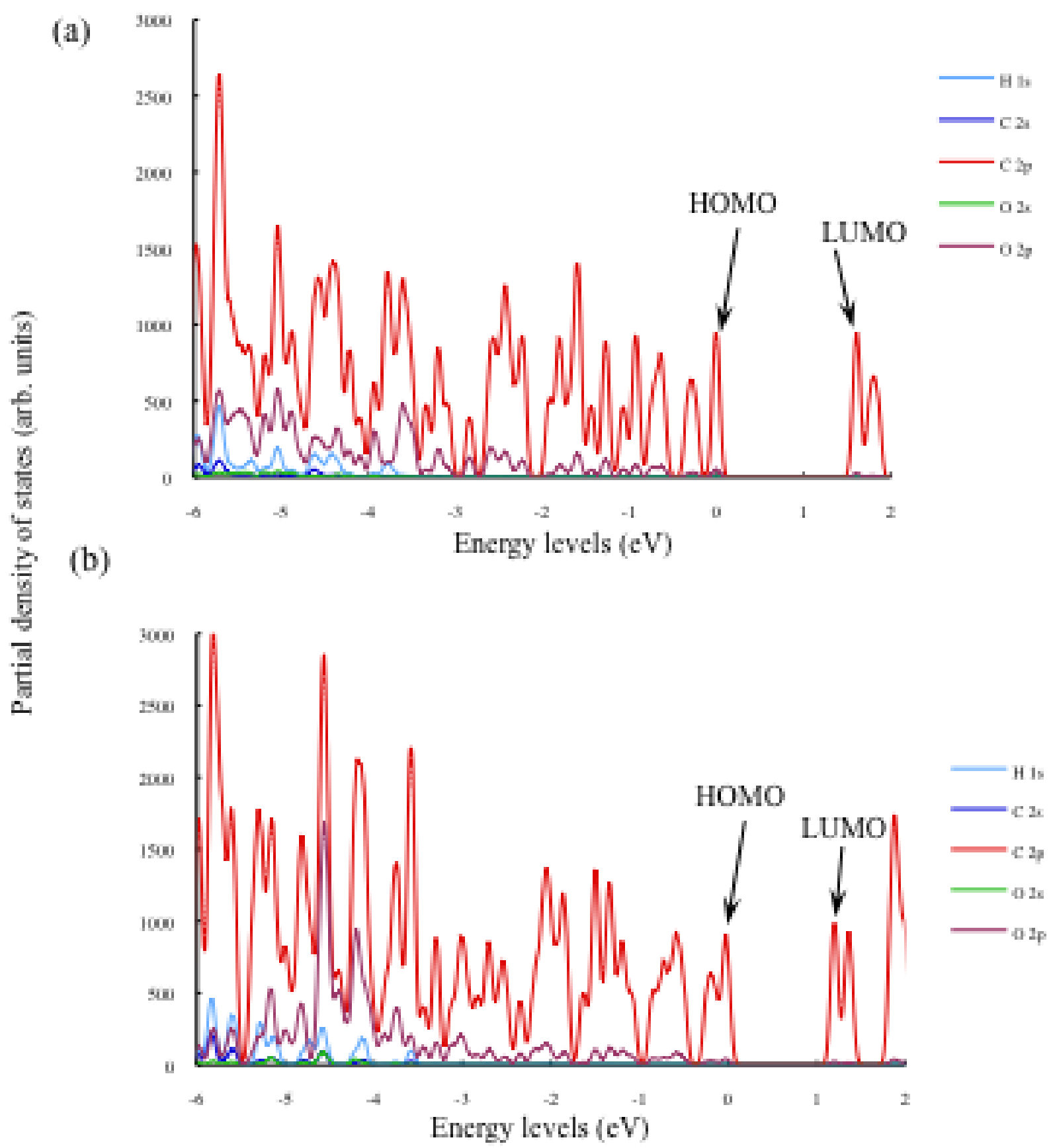

Figure 12 Partial density of states for (a) optimized and (b) unoptimized structure of G(OH) with two ABA stacked functionalized graphene layers.